\documentclass[reprint,aps,prb,amsmath,amssymb]{revtex4-1}

\usepackage[]{graphicx}
\usepackage[]{units}
\usepackage{times}
\usepackage{bm}
\usepackage{ulem}		
\usepackage{color}

\newcommand{\comment}[1]{}

\renewcommand{\emph}{\textit}

\begin{document}

\title{Quenching of Intervalley Exchange Coupling in the Presence of Momentum-Dark States in TMDCs}
\author{Malte Selig$^{1}$}\email{malte.selig@tu-berlin.de}
\author{Florian Katsch$^1$}
\author{Samuel Brem$^2$}
\author{Garnik F. Mkrtchian$^{3}$}
\author{Ermin Malic$^2$}
\author{Andreas Knorr$^1$}
\affiliation{$^1$Nichtlineare Optik und Quantenelektronik, Institut f\"ur Theoretische Physik, Technische Universit\"at Berlin,  10623 Berlin, Germany}
\affiliation{$^2$Chalmers University of Technology, Department of Physics, SE-412 96 Gothenburg, Sweden}
\affiliation{$^3$ Centre of Strong Fields Physics, Yerevan State University, Yerevan, Armenia}

\begin{abstract}
Monolayers of transition metal dichalcogenides are promising materials for valleytronic applications, since they possess two individually addressable excitonic transitions at the non-equivalent $K$ and $K'$ points with different spins, selectively excitable with light of opposite circular polarization. Here, it is of crucial importance to understand the elementary processes determining the lifetime of these optically injected valley excitons. In this study, we perform microscopic calculations based on a Heisenberg equation of motion formalism to investigate the efficiency of the intervalley coupling in the presence (W based TMDCs) and absence (Mo based TMDCs) of energetically low lying momentum-dark exciton states. While we predict a valley exciton lifetime on the order of some hundreds of fs in the absence of low lying momentum-dark states we demonstrate a strong quenching of the valley lifetime in the presence of such states.
\end{abstract}

\maketitle


\section*{Introduction}

\begin{figure*}[t!]
 \begin{center}
\includegraphics[width=1.0\linewidth]{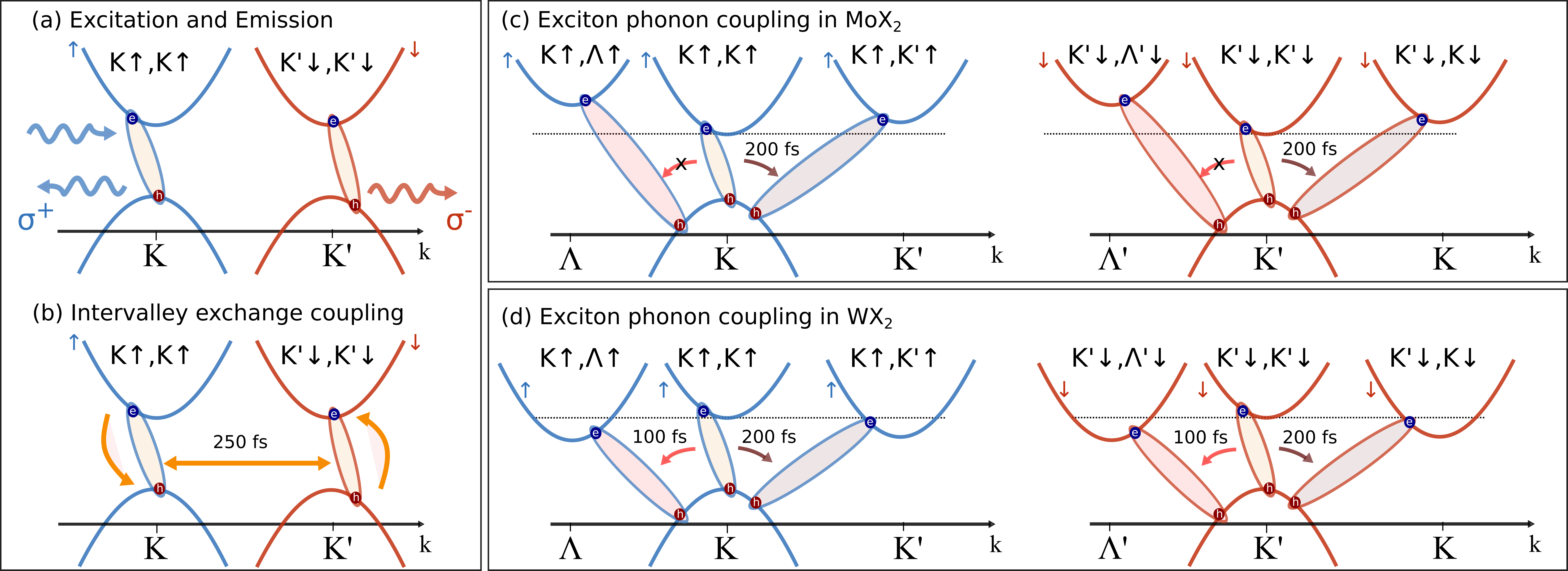}
 \end{center}
 \caption{\textbf{Schematic illustration of the excitonic scattering mechanisms.} Blue bands denote spin up electrons and holes whereas red bands illustrate spin down bands respectively. (a) Excitons at the $K$ ($K'$) valley couple to left (right) handed polarized light. (b) The direct excitons ($(K \uparrow , K \uparrow)$ and $(K' \downarrow , K' \downarrow)$) with same center of mass momentum $\mathbf{Q} = \mathbf{k}_e - \mathbf{k}_h$ are coupled through intervalley exchange coupling. (c) In MoX$_2$ excitons in $(K \uparrow , K\uparrow )$ states can scatter to indirect $(K \uparrow , K' \uparrow)$ states whereas excitons in $(K' \downarrow , K' \downarrow)$ states can scatter to indirect $(K' \downarrow , K \downarrow)$ states. The scattering to $(K \uparrow , \Lambda \uparrow)$ and $(K' \downarrow , \Lambda' \downarrow)$ states is prohibited through the large energetic mismatch. (d) In WX$_2$ excitons in $(K \uparrow , K\uparrow )$ states can scatter to indirect $(K \uparrow , K' \uparrow)$  and $(K \uparrow , \Lambda \uparrow)$ exciton states whereas excitons in $(K' \downarrow , K' \downarrow)$ states can scatter to indirect $(K' \downarrow , K \downarrow)$ and $(K' \downarrow , \Lambda' \downarrow)$ states. At the arrows, indicating the scattering process, we also give approximated effective relaxation rates at \unit[77]{K} which were extracted as $1-e^{-1}$ times from the numerical results. Note, that to the formation of $(K\uparrow,K'\uparrow)$ excitons in WSe$_2$ not only scattering from $(K\uparrow,K\uparrow)$ states, but even more pronounced scattering from $(K\uparrow,\Lambda\uparrow)$ states occurs.}
 \label{schema}
\end{figure*}

Monolayers of transition metal dichalcogenides (TMDCs) possess a variety of excitonic excitations with large binding energy and oscillator strength, which enabled the extensive investigation of exciton physics in these ultrathin materials \cite{Wang2018,Manca2017,Christiansen2017,Steinhoff2017, Steinleitner2018,Kunstmann2018}. These excitons are built up from electron-hole pairs located at the $K$ and the $K'$ point in the first Brillouin zone, being selectively addressable by left (right) handed circularly polarized $\sigma^+$ ($\sigma^-$) light. Due to the spin-valley locking between $K$ and $K'$ point, this allows to create an excitonic spin-valley polarization\cite{Cao2012} (electron and hole spin $\uparrow$ or $\downarrow$), cf. Fig. \ref{schema} (a). Thus, TMDCs are considered to be promising materials for future spin-valleytronic applications\cite{Cao2012} and the valley lifetime describing the intrinsic relaxation time of selectively excited spin-valley excitons is of crucial importance. A complete microscopic understanding of the underlying processes is missing so far. 

The spin polarized valley dynamics was investigated experimentally\cite{Cui2012,Wang2013,Moody2016,Smolenski2016,Schmidt2016,Plechinger2017,McCormick2018} and theoretically\cite{Yu2014,Glazov2014,Wang2014,Wang2014b,Dery2015}: The measured polarizations of the emitted light after circular excitation (spin $\uparrow$, cf. figure \ref{schema} (a)) are typically on the order some ten \% in photoluminescence\cite{Moody2016,Smolenski2016} and the valley lifetimes, i.e. the time which is required to equalize the populations in both valleys, are on the order of hundreds of fs to few ps in pump probe experiments\cite{Wang2013,Conte2015,Schmidt2016,Wang2018,ZWang2018}. As possible candidates for the underlying spin polarization decay via intervalley coupling, pure spin flip mechanisms such as Dyakonov-Perel\cite{Wang2014,Wang2014b}, Elliot Yafet\cite{Wang2014b} mechanisms and the Silva-Sham mechanism\cite{Maialle1993,Vinattieri1994,Glazov2014,Yu2014} have been considered.The latter does not require a single electronic spin flip but flipping both, electron and hole spin at the same time, and is called intervalley exchange coupling (IEC) mechanism in the literature\cite{Qiu2015,Selig2019b}. 
The Dyakonov-Perel mechanism appears in semiconductors without inversion symmetry leading to the formation of an effective magnetic field. In this effective magnetic field the electronic spin precesses which leads to a relaxation of the electron spin.
The Elliot Yafet mechanism appears in materials with strong spin orbit coupling. In such materials, the electronic spin is not a well defined quantum number any more with the result that any scattering event can change the spin. Both spin relaxation mechanisms have been shown to occur on picosecond timescales for in-plane spins\cite{Wang2014,Wang2014b} exceeding the experimentally observed sub picosecond risetime of the nonlinear optical response in the unpumped valley shortly after the optical excitation \cite{Schmidt2016,ZWang2018}. On longer timescales, relevant for the emission from the thermalized excitons, these mechanisms may become important. 
Last, the IEC couples both valleys at the $K$ and $K'$ point through a dipole-dipole interaction flipping both, electron and hole spin at the same time. It was identified as the source of the fast intervalley transfer shortly after the optical excitation \cite{Schmidt2016}. However, since the IEC requires momentum, spin and energy conservation\cite{Glazov2014,Yu2014,Selig2019b} it does not apply for momentum-dark excitons or excitons with opposite spin of the constituent carriers forming the exciton. However, these states can be populated by exciton-phonon scattering and depending on their energetic position (below and above the bright state), they have been demonstrated to be crucial to understand the emission properties of TMDCs \cite{Zhang2015,Selig2018,Lindlau2018,Brem2019b,Glazov2019}. The question arises how thermalization into these momentum-dark exciton states affects the intervalley exchange coupling and the spin polarization.

To answer this question, we develop a theoretical model for IEC based on a Heisenberg equation of motion formalism \cite{Kochbuch,Thranhardt2000,Kira2006,Selig2019b}. 
We account for the fact that the optical properties in TMDCs are determined by tightly bound excitons and introduce an excitonic Hamiltonian including exciton photon, exciton phonon and intervalley Coulomb exchange coupling of excitons describing the system dynamics\cite{Katsch2018}. This Hamiltonian includes the optical accessible states as well as indirect exciton states far beyond the light cone, which were discussed recently\cite{Wu2015,Qiu2015,Selig2016,Selig2018,Niehues2018,Selig2019b}. 
It has been demonstrated that the IEC mechanism is responsible for the ultrafast population of the unpumped valley in the transient regime after optical excitation\cite{Schmidt2016,Selig2019b}, cf. figure \ref{schema} (b). This leads to photoluminescence from the unpumped valley on a sub-picosecond timescale. However, as we will discuss in the present study, we find significant differences in the intervalley relaxation between molybdenum and tungsten bases TMDCs. While in molybdenum based TMDCs, shortly after optical excitation of the $K$ valley, cf. figure \ref{schema} (a), most excitons occupy $(K \uparrow , K \uparrow)$ states, and only a small amount of excitons scatter to $(K \uparrow , K' \uparrow)$ states, since these states are located slightly above the radiative cone, cf. figure \ref{schema} (c). The simultaneously occuring intervalley exchange coupling between $(K \uparrow , K \uparrow)$ and $(K' \downarrow , K' \downarrow)$ states, cf. figure \ref{schema} (b), is very efficient since most of the excitons will thermalize in those states. In contrast, in tungsten based TMDCs, phonon mediated thermalization causes the relaxation of excitons to energetically low lying $(K \uparrow , \Lambda \uparrow)$  and $(K \uparrow , K' \uparrow)$, cf. figure \ref{schema} (d). These states are inactive for the intervalley exchange coupling, since the required energy and momentum selection rules can not be fulfilled at the same time here. The thermalization into these states drastically reduces the amount of excitons in the $(K \uparrow , K \uparrow)$ states, and as a results the intervalley exchange coupling efficiency is quenched. In particular the calculated timescales for the IEC exceed the timescales which were predicted for the spin relaxation of the Dyakonov-Perel and Elliot Yafet type \cite{Wang2014,Wang2014b}.

\section*{Theoretical Approach}


First we introduce an excitonic Hamiltonian including exciton-photon, exciton-phonon and exciton-exciton Coulomb exchange coupling. We restrixt the description of excitonic correlations to the second order of the exciting field\cite{Thranhardt2000}. This Hamiltonian is discussed in detail in the supplementary material. To access the excitonic wavefunction and energy dispersion, we exploit the Wannier equation, where the used Coulomb potential takes into account the dielectric environment and is treated beyond the Rytova-Keldysh limit\cite{Trolle2017}, taking the full momentum dependence of the dielectric function into account, which allows for a good approximation of screened Coulomb interaction obtained from DFT calculations\cite{Latini2015,Qiu2016,Steinhoff2017}. The parameters for the underlying electronic bandstructure and electron phonon couplings are taken from DFT calculations \cite{Kormanyos2015,Li2013,Jin2014}. To derive equations of motion from this Hamiltonian, we exploit Heisenbergs equation of motion. In the following we will discuss the required equations of motion in detail.

To discuss the photoemission dynamics, we calculate the photon number $n^\sigma_{\mathbf{K}K_z}$ of the emitted light with the three dimensional wave vector $(\mathbf{K},K_z)$, with $\mathbf{K}$ denoting the two-dimensional component within the semiconductor plane and $K_z$ its perpendicular component, and the polarization $\sigma = \sigma^+ , \sigma^-$ being defined as $n^{\sigma}_{\mathbf{K},K_z}=\langle c_{\mathbf{K},K_z}^{\dagger \sigma} c_{\mathbf{K},K_z}^{\sigma} \rangle$ via $c^{(\dagger) \sigma}_{\mathbf{K},K_z}$ denoting annihilation (creation) operators for photons\cite{Loudon}. The total photoluminescence intensity for a certain light polarization is found by summing the photon rate over all momenta $(\mathbf{K},K_z)$, i.e.  $I^{\sigma}\propto \sum_{\mathbf{K},K_z} \omega_{\mathbf{K},K_z}\partial_t n_{\mathbf{K},K_z}^{\sigma}$, with the photon frequency $\omega_{\mathbf{K},K_z}$\cite{Thranhardt2000}, which, in Born Markov approximation reads\cite{Kochbuch,Thranhardt2000,Selig2018} 
\begin{align}
I^{\sigma}\propto\frac{2 \pi}{\hbar}\hspace{-2pt}\sum_{{\mathbf{K},K_z},\xi} |d^{\xi \sigma}_{\mathbf{K}}|^2 \left(  |\langle P_\mathbf{K}^{\xi\xi} \rangle|^2 +  N_\mathbf{K}^{\xi\xi}\right)\delta( \Delta E^{\xi \sigma}_{\mathbf{K},K_z} ).\label{eqPL}
\end{align}
$I^\sigma$ is proportional to the amount of coherent ($|\langle P_\mathbf{K}^{\xi\xi} \rangle|^2$) and incoherent exciton density ($N_\mathbf{K}^{\xi\xi}$) within the light cone $|\mathbf{K}|\leq K_L = \frac{\omega^{1s}}{c}$ resulting from the delta function $\Delta E^{\xi \sigma}_{\mathbf{K},K_z}=E^{\xi}_\mathbf{K}-\hbar \omega^{\sigma}_{\mathbf{K},K_z}$ which ensures the energy conservation during the photon emission. Here, $E^{\xi}_\mathbf{K}$ denotes the excitonic energy. $\mathbf{d}^{\xi \sigma}_\mathbf{K}$ denotes the excitonic dipole matrix element (circular polarization $\sigma$ and valley $\xi$).
The incoherent exciton density is denoted by $ N_\mathbf{Q}^{\xi_h \xi_e}=\delta \langle P^{\dagger \xi_h \xi_e}_\mathbf{Q} P^{\xi_h \xi_e}_\mathbf{Q} \rangle$, where we have introduced excitonic operators $P^{\xi_h\xi_e}_{\mu,\mathbf{Q}}$ \cite{HaugIvanov1993,Katsch2018,Lengers2019}, cp. Eq. S2 in the supplementary material, with the merged valley $i_{h/e}$ and spin $s_{h/e}$ index for holes and electron $\xi_{h/e}  = (i_{h/e},s_{h/e})$, the excitonic state quantum number $\mu$ and the two-dimensional center of mass momentum $\mathbf{Q}$. Additionally, $\delta \langle .. \rangle$ accounts for the purely correlated part of the expectation value\cite{Thranhardt2000}. For the underlying electronic bandstructure we include the high symmetry points $i_e \in \{ K,K',\Lambda,\Lambda'\}$ and for holes we include the high symmetry points $i_h \in \{K,K'\}$ explicitly to our investigation, cf. figure \ref{schema} (a) and (b). In the following we restrict our analysis to the lowest bound exciton state $\mu=1s$ justified by the large energy difference between $1s$ and $2s$ exciton states \cite{Chernikov2014,Brem2018,Brem2019a}. 
To calculate the coherent emission from Eq. \ref{eqPL}, we derive an equation of motion for the excitonic coherence $\langle P^{\xi_h \xi_e}_\mathbf{Q} \rangle$

\begin{align}
i \hbar \partial_t \langle P^{\xi_h \xi_e}_\mathbf{Q} \rangle&= (E_\mathbf{Q}^{\xi_h \xi_e}-i \gamma^{\xi_h \xi_e}_\mathbf{Q}) \langle P^{\xi_h \xi_e}_\mathbf{Q}\rangle \nonumber \\&+ \sum_{\sigma} \mathbf{d}^{\xi_h\sigma}\cdot \mathbf{E}^\sigma \delta^{\xi_h,\xi_e}_\mathbf{Q,0}
+X^{\xi_h \bar{\xi_h}}_\mathbf{Q} \langle P^{\bar{\xi_h} \bar{\xi_e}}_\mathbf{Q}\rangle\delta_{\xi_h,\xi_e}.\label{CoherenceEq}
\end{align}
The first line describes the excitonic dispersion with the excitonic energy $E_\mathbf{Q}^{\xi_h\xi_e}$ and includes the dephasing of the excitonic coherence $\gamma_\mathbf{Q}^{\xi_h\xi_e}$ consistently calculated from radiative coupling and exciton phonon scattering\cite{Selig2016,Christiansen2017,Brem2019a}. The first term in the second line in Eq. \ref{CoherenceEq} represents the optical excitation of the excitonic coherence with a coherent light pulse $\mathbf{E}^\sigma(t)$. The second term in the second line describes the intervalley Coulomb exchange coupling of the excitonic coherences between $K$ and $K'$ valley, which is characterized by the matrix element $X^{\xi_h \bar{\xi_h}}_\mathbf{Q}$, cf. the supplementary material. Troughout this paper, we assume weak excitation, such that the hierachy problem which arises from the Coulomb interaction can be truncated at the lowest order\cite{Axt1994a,Kira2006}. Therefore Pauli blocking and other many body effects do not occur in Eq. \ref{CoherenceEq}. Since the coupling element $X^{\xi_h \bar{\xi_h}}_\mathbf{Q}$ is proportional to $|\mathbf{Q}|$ in the lowst order \cite{Qiu2015,Wu2015,Schmidt2016} and the optically injected excitonic coherences have vanishing center of mass momentum (only $P_\mathbf{Q=0}$ is excited by the incident light) the excitonic coherence in one valley $\xi$ does not influence the coherence in the opposite valley $\bar{\xi}=(\bar{i},\bar{s})$ (i.e. $\xi=(K, \uparrow)$ and $\bar{\xi} = (K',\downarrow)$).
This does not apply for the momentum-dependent incoherent exciton density $N_\mathbf{Q}$ in the $K$, since due to scattering finite $\mathbf{Q}$'s are occupied. Its equation of motion reads
\begin{align}
\partial_t N_\mathbf{Q}^{\xi_h\xi_e}&= \Gamma^{in\,\xi_e-\xi_h}_\mathbf{Q} |\langle P^{\xi_h\xi_h}_\mathbf{0}\rangle|^2  \nonumber \\ &+\sum_{\mathbf{K},\xi_e'} \Gamma_\mathbf{Q,K}^{in\,\xi_e-\xi_e'} N^{\xi_h\xi_e'}_\mathbf{K} - \sum_{\mathbf{K},\xi_e'} \Gamma_\mathbf{Q,K}^{out\,\xi_e - \xi_e'} N^{\xi_h \xi_e}_\mathbf{Q} \nonumber \\&- \Gamma^{\xi_h-\xi_e}_\mathbf{Q} N^{\xi_h\xi_e}_\mathbf{Q}\nonumber\\&+\frac{2}{\hbar} \Im{(X^{\xi_h \bar{\xi_h}}_\mathbf{Q} C^{\xi_h \bar{\xi_h}}_\mathbf{Q})(\delta_{\xi_h,K\uparrow}^{\xi_e,K\uparrow}-\delta_{\xi_h,K'\downarrow}^{\xi_e,K'\downarrow})}.\label{density}
\end{align}
The first line accounts for the exciton-phonon scattering mediated formation of incoherent excitons driven by the optically excited excitonic coherence $P^{\xi_h\xi_h}_\mathbf{0}$\cite{Thranhardt2000,Selig2018,Brem2018}, compare Eq. \ref{CoherenceEq}. The notation $(\xi_h,\xi_h)$ implies that the excitonic coherence can only be optically addressed in momentum-bright ($\mathbf{Q} = 0$) exciton states, cp. Eq. \ref{CoherenceEq}. The second line can be identified as a Boltzmann scattering equation accounting for the thermalization of the incoherent exciton densities and cooling into a Boltzmann distribution. This includes exciton-phonon sacttering within the excitonic valleys as well as between them, cf. Fig. \ref{schema} (c) and (d).
The exciton-phonon scattering rates $\Gamma_\mathbf{Q,K}^{in\,\xi_e \xi_e'}$ and $\Gamma_\mathbf{Q,K}^{out\,\xi_e \xi_e'}$ are defined in the appendix, cp. Eqs. S13 and S14 in the supplementary material.
The third line of Eq. \ref{density} describes the radiative decay of the exciton density with the momentum-dependent relaxation rate $\Gamma^{\xi_i\xi_j}_\mathbf{Q}=\frac{2}{\hbar}\gamma_{\text{rad}} \sum_\mathbf{K}\delta( \Delta E^{\xi \sigma}_\mathbf{K} )\delta_\mathbf{Q,K_\parallel}$, cf. Fig. \ref{schema} (a). The appearing Kronecker $\delta$ ensures the energy conservation during the photon emission and accounts for the fact that only excitons which are located within the light cone can decay radiatively. As a result, electron and hole which form the exciton, have to be located at the same high symmetry point in the 1. Brillouin zone to decay radiatively. So far, the discussed contributions account for the formation, thermalization and photoluminescence of excitons.

However, the last term in Eq. \ref{density} is the most important one for our work and describes the  intervalley Coulomb exchange dynamics responsible for the intervalley excitation transfer, cf. Fig. \ref{schema} (b). It acts as a exchange for exciton densities in $(K \uparrow , K \uparrow)$ and $(K' \downarrow , K' \downarrow)$ states conserving the total amount of excitons. 
This interaction is mediated by the intervalley coherence $C^{\xi \bar{\xi}}_\mathbf{Q}$ between the $(K\uparrow, K\uparrow )$ and the $(K'\downarrow , K'\downarrow)$ states in Eq. \ref{density} defined as $ C_\mathbf{Q}^{\xi \bar{\xi}}=\langle P^{\dagger \xi \xi}_\mathbf{Q} P^{\bar{\xi}\bar{\xi}}_\mathbf{Q} \rangle$. The equation of motion for the intervalley coherence reads
\begin{equation}
\partial_t C_\mathbf{Q}^{\xi \bar{\xi}}=\partial_t C_\mathbf{Q}^{\xi \bar{\xi}} \Big \vert_{scat}+\frac{1}{i \hbar} X^{\bar{\xi} \xi}_\mathbf{Q} (N^{\xi \xi}_\mathbf{Q}-N^{\bar{\xi} \bar{\xi}}_\mathbf{Q}).\label{coherence}
\end{equation}
The first term in Eq. \ref{coherence} i.e. $\partial_t C_\mathbf{Q}^{\xi \bar{\xi}}\Big \vert_{scat}$, describes the exciton-phonon interaction and leads to both diagonal and off-diagonal dephasing of the intervalley coherence, cf. the supplementary material. The second term in Eq. (\ref{coherence}) acts as a source for the intervalley coherence driven by the occupation difference of the exciton densities in the opposite valleys, i.e. $(N^{\uparrow}_\mathbf{Q}-N^{\downarrow}_\mathbf{Q})$. 

At this point, we want to highlight, that due to our calculations momentum forbidden exciton states do not contribute to the intervalley exchange coupling, since it requires energy and momentum conservation, which can only be fulfilled by $(K,K)$ and $(K'.K')$ excitons.
Also, in our analysis, we do not include coupling mechanisms which lead to a spin flip of an individual carrier and address the investigation of these processes to future work. Since for optically excited excitons, the electron and hole spin coincide, we can conclude, that all excitons in our analysis have the same electron and hole spins.


\section*{Results}

Numerically evaluating Eqs. \ref{eqPL} - \ref{coherence}, we have microscopic access to the time- and momentum-resolved intervalley dynamics, including the temporal evolution of the optically injected excitonic coherence in the $(K\uparrow,K\uparrow)$ valley, the valley resolved excitonic occupations $N^{K\uparrow,i_e \uparrow}=\sum_{\mathbf{Q}} N_\mathbf{Q}^{K \uparrow i_e \uparrow}$ and  $N^{K'\downarrow , i_e \downarrow}=\sum_{\mathbf{Q}} N_\mathbf{Q}^{K' \downarrow   i_e \downarrow}$, respectively, cp. Fig \ref{schema}, as well as the photoemission intensity $I^\sigma$ with respect to the polarization $\sigma$. As exemplary materials we investigate  MoSe$_2$  and WSe$_2$ on a SiO$_2$ substrate. Due to our evaluation of the Wannier equation\cite{Selig2018}, in MoSe$_2$ the bright $(K\uparrow,K\uparrow)$ and $(K'\downarrow,K'\downarrow)$ states are the energetically lowest state in the excitonic Brillouin zone being located a few meV below the $(K\uparrow,K'\uparrow)$ and $(K'\downarrow,K\downarrow)$ states, cf. Fig. \ref{schema} (c), whereas in WSe$_2$ the momentum-dark $(K\uparrow,\Lambda\uparrow)$, $(K'\downarrow,\Lambda'\downarrow)$, $(K\uparrow,K'\uparrow)$ and $(K'\downarrow,K\downarrow)$ states are located energetically below the optical bright state by some tens of meV, cf. Fig. \ref{schema} (d)\cite{Selig2016,Malic2018}. 

By comparing both cases, we are able to investigate the influence of energetically low lying momentum-dark states to the intervalley exchange coupling dynamics. We want to note, that the exact quantitative position of these momentum-dark states is still under debate in the literature\cite{Deilmann2019}.
\subsection*{Intervalley coupling in MoSe$_2$}
\begin{figure}[t!]
 \begin{center}
\includegraphics[width=1.0\linewidth]{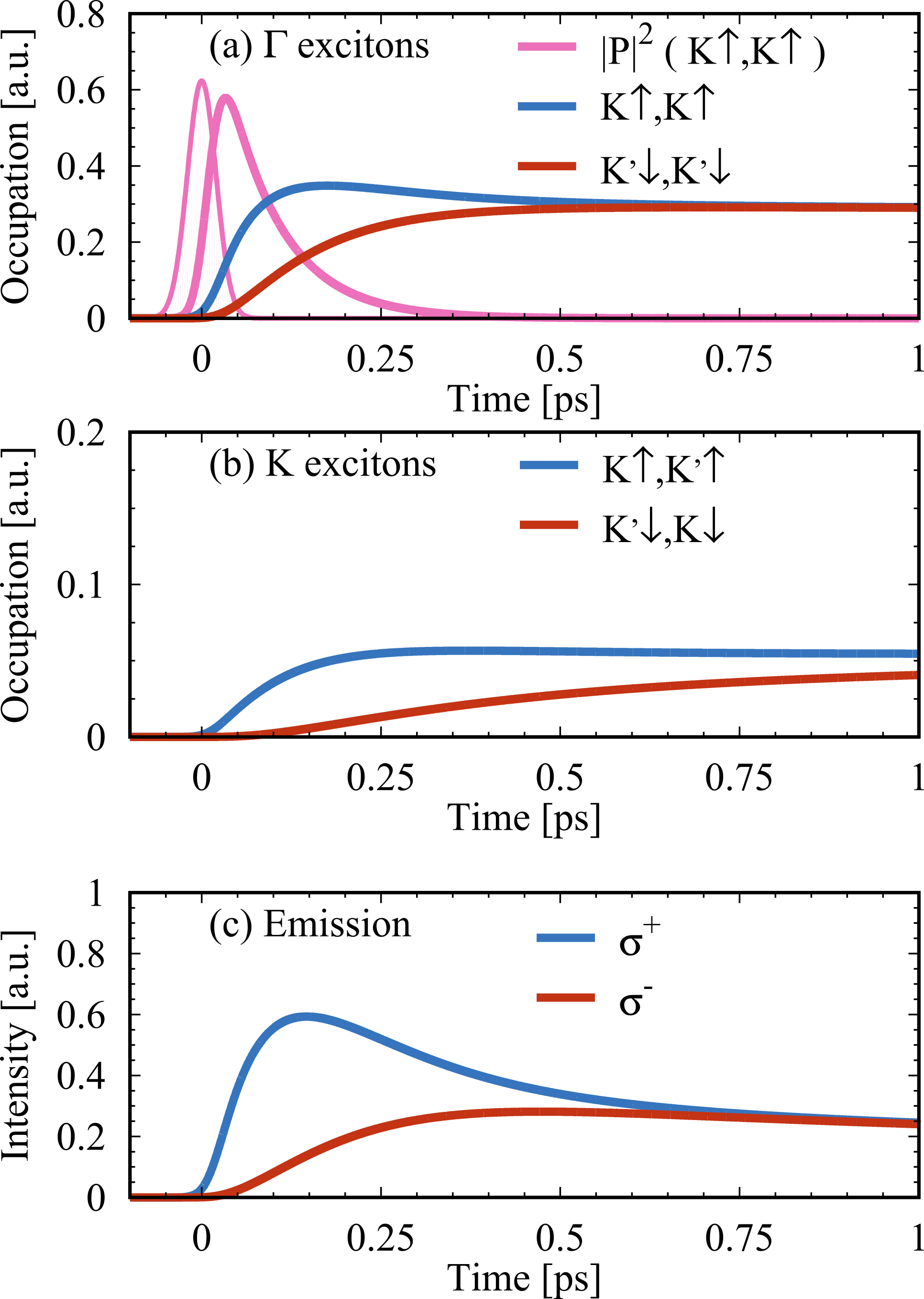}
 \end{center}
 \caption{\textbf{Time evolution of exciton density and intensity of photoemission at \unit[77]{K} in MoSe$_2$.} Figure (a) illustrates the temporal evolution of the $\Gamma$ excitons. Additionally we show the light pulse (pink shaded) as well as the optically injected excitonic transition (pink solid). Figure (b) shows the temporal evolution of the $K$ excitons.  In figure (d) we show the corresponding emission intesities of $\sigma^+$ and $\sigma^-$ light.}
 \label{DepolMoSe2}
\end{figure}

Figure \ref{DepolMoSe2} illustrates the intervalley dynamics in MoSe$_2$ at an exemplary temperature of \unit[77]{K} after optically exciting the $(K\uparrow,K\uparrow)$ valley ($A$ states) resonantly to the 1s transition with a left handed polarized $\sigma^{+}$ \unit[20]{fs} gaussian light pulse, cf. figure \ref{DepolMoSe2} (a). The optically excited excitonic coherence decays due to radiative and exciton-phonon interaction within \unit[300]{fs} being consistent with previous calculations\cite{Selig2018}. Due to the non-radiative decay of the excitonic coherence through exciton phonon scattering, incoherent excitons with non-vanishing center of mass momenta are formed in the $(K\uparrow,K\uparrow)$ states on a similar timescale. From these states, incoherent exciton in $(K'\downarrow,K'\downarrow)$ states are formed through Coulomb intervalley exchange coupling, cf. Fig. \ref{schema} (b). We find that both densities equilibrate after approximately \unit[500]{fs}.

In Fig. \ref{DepolMoSe2} (b), we show the density of the indirect $(K \uparrow , K' \uparrow )$ and $(K' \downarrow , K \downarrow )$ excitons. Due to our calculations, these states are located \unit[10]{meV} above the $(K\uparrow , K\uparrow )$ and $(K' \downarrow , K' \downarrow)$ states respectively (bright ground state in MoSe$_2$).  We note, that recent DFT calculations predicted another value of the energetic separation of the direct $(K,K)$ and indirect $(K,K')$ states\cite{Deilmann2019}. However, the deviations are only on the order of few meV and smaller in comparison to the thermal energy of the excitons. Such deviations however, do not induce significant changes in the exciton dynamics for example the opening of new relaxation channels. 

In Fig. \ref{DepolMoSe2} (b), we find the formation rate of the $(K\uparrow , K' \uparrow )$ excitons in the order of \unit[300]{fs} due to exciton phonon scattering in expense to the optically excited $(K \uparrow , K \uparrow)$ exciton, being consistent with previous work\cite{Selig2018}. In contrast, the formation time of the $(K'\downarrow , K \downarrow )$ exciton occurs much slower within \unit[2]{ps}. This is due to the fact, that at least 3 scattering events are needed to bring excitons to these states: (i) finite wavenumber excitons within the $(K \uparrow , K \uparrow)$ valley have to be created through phonon scattering from the optically injected excitonic coherence to switch on the exchange interaction: (ii) intervalley coupling leads then to the formation of $(K' \downarrow , K' \downarrow)$ exctions, (iii) and finally exciton phonon scattering results in excitons in $(K'\downarrow , K \downarrow )$ states.

Fig. \ref{DepolMoSe2} (c) shows the temporal evolution of the emitted light intensity. The $\sigma^{+}$ polarized emission from the $( K \uparrow , K \uparrow )$ valley  starts directly after the optical exciation, since it results from the optically injected exciton density. In contrast, the $\sigma^{-}$ polarized emission exhibits a delay since this emission stems from the $N^{K'\downarrow K'\downarrow}$ exciton density which first has to be created via intervalley exchange coupling. Additionally we find, that the $\sigma^-$ emission from the unpumped valley is also delayed with respect to the exciton density dynamics in the unpumped valley. The reason is, that excitons which were created through intervalley exchange coupling in these states have non-vanishing center of mass momenta (due to the linear $|\mathbf{Q}|$ dependence of the IEC\cite{Qiu2015,Wu2015}) and therefore have to relax to momentum states within the light cone by phonon scattering. The difference between this times is given by the time, which excitons need to scatter down from states with elevated energies into the lightcone.

\subsection*{Intervalley coupling in WSe$_2$}

\begin{figure}[t!]
 \begin{center}
\includegraphics[width=1.0\linewidth]{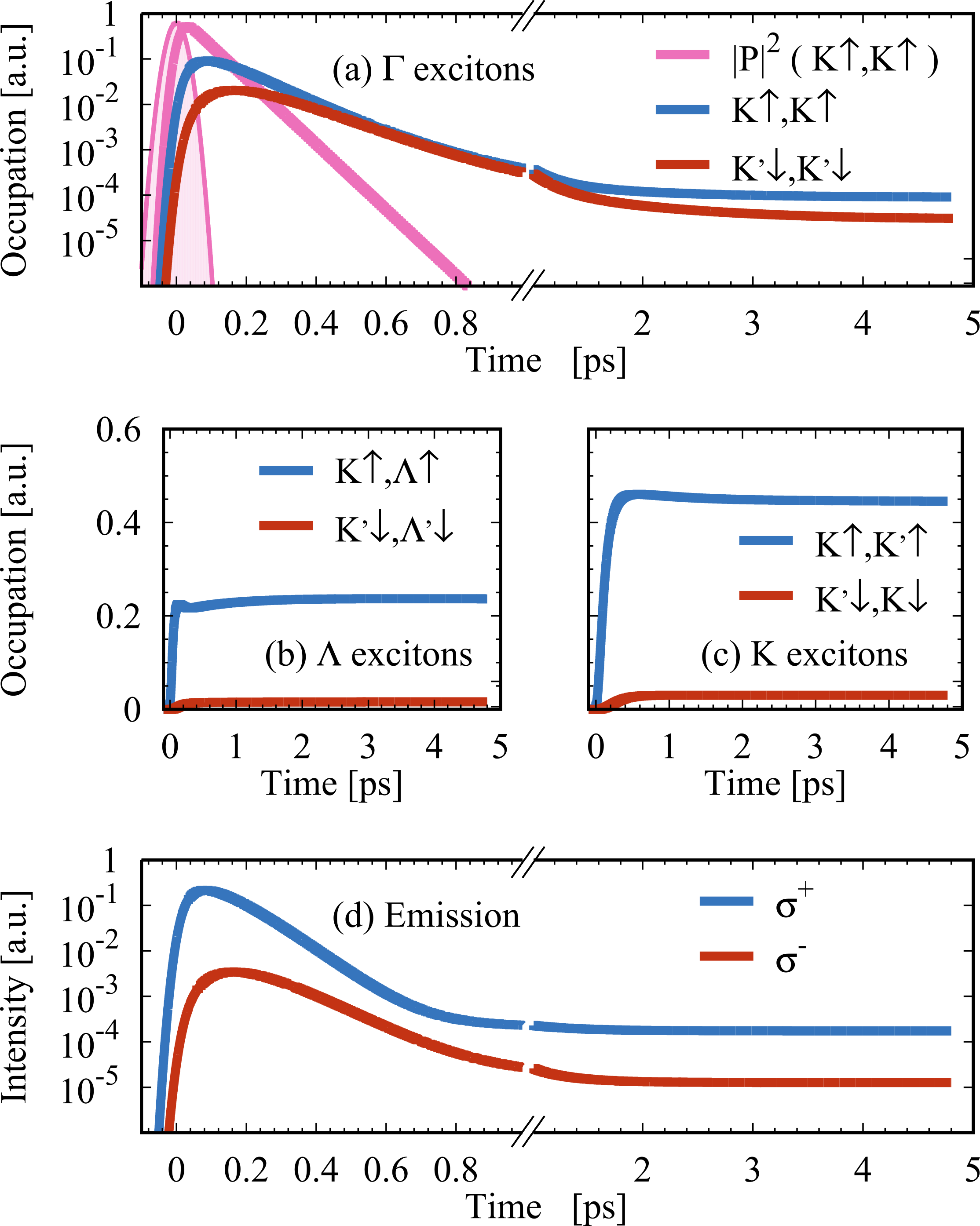}
 \end{center}
 \caption{\textbf{Time evolution of exciton density and intensity of photoemission at \unit[77]{K} in WSe$_2$.} (a) shows the time evolution of the $\uparrow$ and $\downarrow$ $\Gamma$ exciton densities (note the logscale), (b) and (c) illustrate the corresponding $\Lambda$ and $K$ exciton densities. In Fig. (b) we show the corresponding emission intesities of $\sigma^+$ and $\sigma^-$ light. }
 \label{DepolWSe2}
\end{figure}

We now investigate the spin valley dynamics in monolayer WSe$_2$, where momentum-dark $(K,K')$ and $(K,\Lambda)$ states are located energetically below the optically bright state \cite{Selig2016,Selig2018,Malic2018}, cf. Fig \ref{schema} (d). 
Fig. \ref{DepolWSe2} exhibits the spin and valley resolved exciton densities as well as the polarization resolved photoluminescence intensity as a function of time.
In Fig. \ref{DepolWSe2} (a) we depict the time dynamics of the exciton density in $(K \uparrow , K \uparrow)$ and $(K' \downarrow , K' \downarrow)$ states. As in the case of MoSe$_2$, the $(K \uparrow , K \uparrow)$ density is created from the optically pumped exciton coherence through exciton phonon scattering on the timescale of the coherence lifetime and exhibits a subsequent decay through relaxation to low lying momentum-dark $(K \uparrow , \Lambda \uparrow )$ and $(K\uparrow , K' \uparrow)$ states. As long as there is sufficient density in the $(K \uparrow , K \uparrow)$ states, intervalley exchange coupling transfers excitons to the $(K' \downarrow , K' \downarrow)$ states. However since the phonon mediated relaxation to the momentum-dark and energetically low lying $(K \uparrow,\Lambda \uparrow)$ and $(K \uparrow,K \uparrow)$  is comparably fast, the $(K' \downarrow , K' \downarrow)$ exciton density increases only weakly during the first ps after the pump and is not substantially populated. Accordingly, the residual occupation difference between both states decreases on a ns timescale. The reason for this long timescale (in comparison to MoSe$_2$) will be discussed in the following:

In Fig. \ref{DepolWSe2} (b) we show the time evolution of the $( K\uparrow , \Lambda \uparrow )$ and the $(K' \downarrow , \Lambda' \downarrow)$ exciton. Due to the energetic structure, the $( K\uparrow , \Lambda \uparrow )$ excitons can be formed by efficient intervalley phonon scattering through phonon emission from $(K \uparrow , K \uparrow)$ excitons. We find a formation rate of approximately \unit[100]{fs}, consistent with previous studies \cite{Selig2018}. In contrast, the formation rate of the $(K' \downarrow , \Lambda' \downarrow )$ excitons is delayed by some hundreds of fs, since these excitons can only be formed by exciton phonon scattering from $(K' \downarrow , K' \downarrow)$ excitons, which first have to be formed through intervalley exchange coupling from the $(K \uparrow , K \uparrow)$ excitons, cf. Fig. \ref{schema} (a). For the $(K\uparrow , K' \uparrow )$ and $(K'\downarrow , K \downarrow )$ exciton densities, cf. Fig. \ref{DepolWSe2} (c), we find the same qualitative behavior as for the $( K \uparrow , \Lambda \uparrow )$ and $(K' \downarrow , \Lambda' \downarrow)$ excitons, but due to a less efficient exciton phonon scattering with $K$ phonons compared to $\Lambda$ phonons\cite{Li2013,Jin2014}, slightly longer timescales. The formation of the $( K \uparrow K'\uparrow )$ exictons occurs within \unit[200]{fs} while the formation of $(K'\downarrow , K \downarrow )$ excitons takes place within \unit[500]{fs}.
We find that after \unit[300]{fs}, most excitons are located in low lying momentum-dark $(K \uparrow, \Lambda \uparrow )$ and $(K \uparrow , K' \uparrow)$ states. \textit{Since the intervalley exchange coupling does not occur for these indirect excitons, this blocks the intervalley exchange spin relaxation. This results in a pronounced occupation difference between the $(K \uparrow , K' \uparrow)$ and $( K'\downarrow , K \downarrow )$ as well as $(K \uparrow , \Lambda \uparrow)$ and $( K'\downarrow , \Lambda' \downarrow )$ exciton states which persists on a ns timescale.}

In Fig. \ref{DepolWSe2} (d), we show the time dependence of the polarization resolved emission. We find for the $\sigma^+$ polarized light an ultra fast increase, since it stems from the $N^{(K \uparrow , K \uparrow)}$ density, which is initialized optically. The fast rise in the intensity is followed by a decay of \unit[2]{ps} which we address to phonon mediated relaxation to low lying $(K\uparrow, \Lambda \uparrow )$ and $(K \uparrow, K' \uparrow )$ states\cite{Selig2018}. The emission of $\sigma^-$ polarized light starts with a delay compared to the corresponding exciton occupation, similar as in MoSe$_2$, cf. Fig. \ref{DepolMoSe2} (c). Again, this is due to the fact that the formation of excitons within the light cone of the $(K'\downarrow, K'\downarrow)$ states requires at least three scattering events, compare the discussion for MoSe$_2$. Similar to the $\sigma^+$ emission, we again find a decay of \unit[2]{ps} for the $\sigma^-$ emission, which we again attribute to the relaxation to $(K' \downarrow , \Lambda' \downarrow)$ and $(K' \downarrow , K \downarrow )$ states.
After this time, both, the $\sigma^+$ and $\sigma^-$ emission decay on a long timescale, which is determined by radiative decay\cite{Selig2018}. Again, the residual difference in the intensities stays on a ns timescale.

\subsection*{Degree of Polarization and Valley Lifetime}

\begin{figure}[t!]
 \begin{center}
\includegraphics[width=0.9\linewidth]{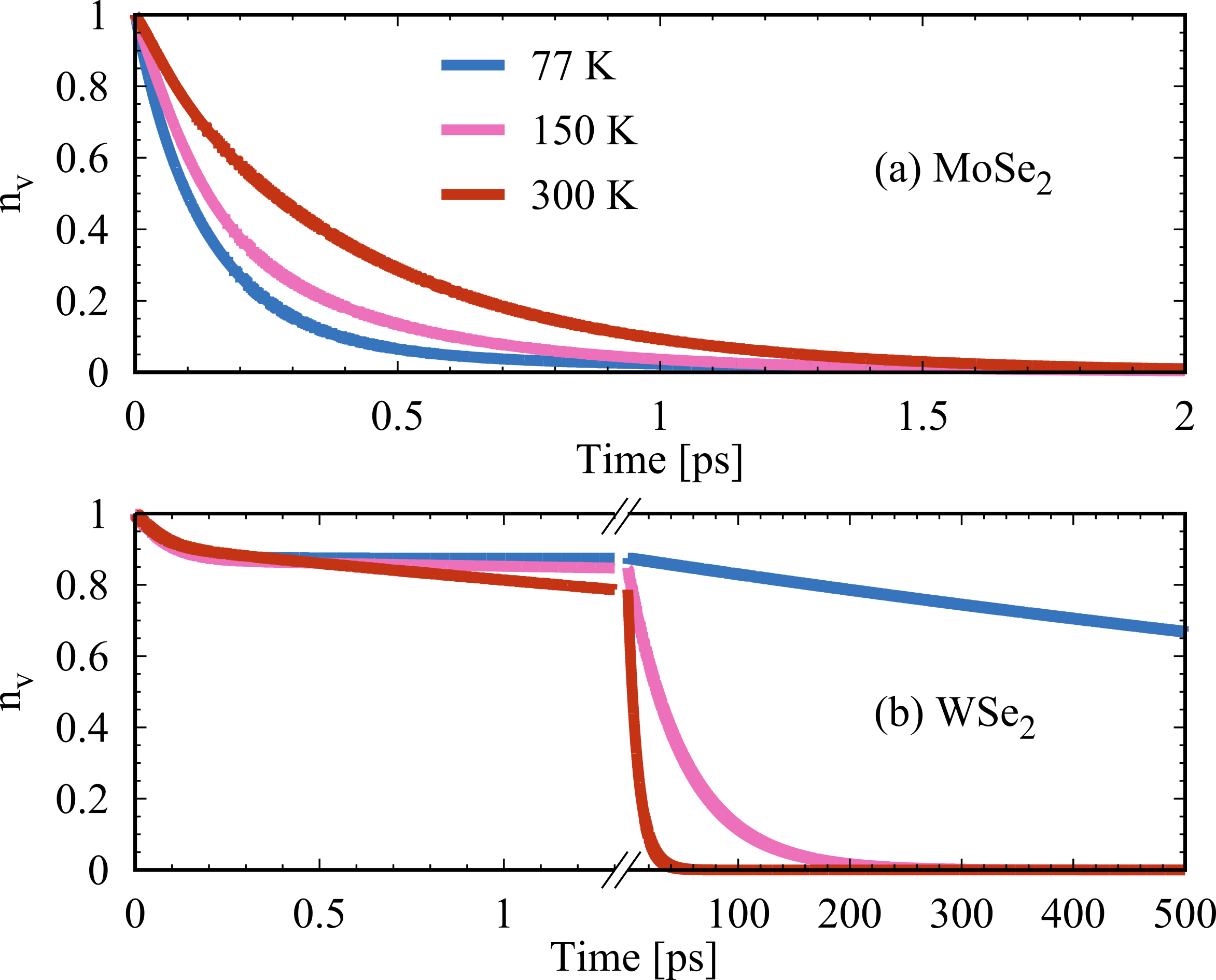}
 \end{center}
 \caption{\textbf{Relaxation dynamics} We show the normalized difference in the occupation $n_v=\frac{N^{K \uparrow}-N^{K \downarrow}}{N^{K \uparrow}+N^{K \downarrow}}$ as a function of time after optically excitation at different temperatures in MoSe$_2$ (a) and WSe$_2$ (b)}
 \label{Difference}
\end{figure}

The intervalley dynamics in both materials, so far discussed at \unit[77]{K}, can be evaluated at various temperatures and used to discuss the degree of the valley polarization lifetime and the polarization of the emission. In Fig. \ref{Difference} , we show the normalized occupation difference of both spin population $n_v=\frac{N^{\uparrow \uparrow}-N^{\downarrow \downarrow}}{N^{\uparrow \uparrow}+N^{\downarrow \downarrow}}$, with $N^{ss} = \sum_{\mathbf{Q},i_h,i_e} N^{i_h s i_e s}_\mathbf{Q}$, in MoSe$_2$ (a) and WSe$_2$ (b). The valley lifetime is defined as the time, after which $n_v$ reaches a value of $e^{-1}$. As discussed before, at \unit[77]{K}, we find a valley lifetime of the exciton density of approximately \unit[200]{fs} in MoSe$_2$. An increase of the temperature leads to an increase of the valley time to \unit[800]{fs} at room temperature. This is due to the fact, that at elevated temperatures more and more excitons are located in the $(K\uparrow, K'\uparrow )$ states and do not contribute to the intervalley exchange coupling. Thus the spin polarization decay becomes less efficient and the valley lifetime enlarges.

In WSe$_2$, cf. Fig. \ref{Difference} (b), we observe a biexponential decay of $n_v$ at all investigated temperatures. Here, the first decay can be ascribed to the interplay of phonon mediated exciton thermalization and intervalley exchange coupling in the transient, non-thermal regime shortly after the optical excitation. We find that this first decay becomes less pronounced at elevated temperatures: At low temperature, excitons are created from coherent excitons in the $(K\uparrow , K\uparrow )$ as well as in the momentum-dark states through phonon scattering. Shortly after the excitation, we have much more excitons in the $(K\uparrow , K\uparrow )$ compared to the situation after thermalization. These excitons can also couple to the $(K' \downarrow , K' \downarrow)$ valley via intervalley exchange coupling leading to the initial decay of $n_v$. At elevated temperatures, exciton phonon coupling becomes more efficient resulting in a faster thermalization. 

For the long time component we observe an inverse behavior. As already discussed, at \unit[77]{K}, we find a ultralong lifetime of some ns, since most excitons are located in indirect momentum-dark states, which efficiently suppresses the intervalley exchange interaction. At elavated temperatures, also higher energy states are occupied, especially $(K \uparrow , K \uparrow)$ which then can couple to $(K' \downarrow , K' \downarrow)$ states via intervalley exchange coupling. Thus we observe a valley lifetime time of \unit[10]{ps} at room temperature.

\begin{figure}[t!]
 \begin{center}
\includegraphics[width=0.9\linewidth]{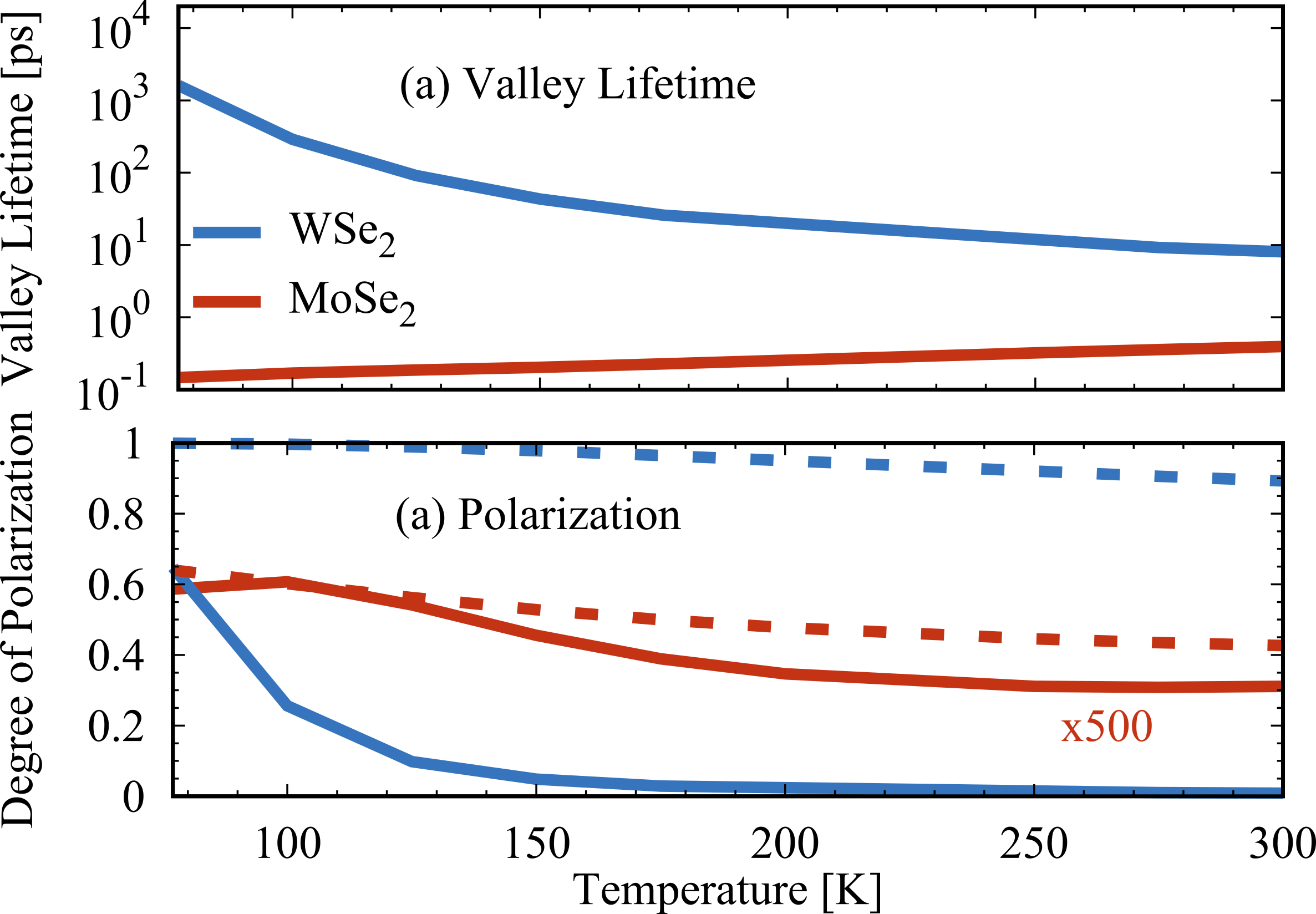}
 \end{center}
 \caption{\textbf{Polarization degree and Valley Lifetime} (a) Valley lifetime in both investigated materials, extracted from the data in figure \ref{Difference}. (b) Degree of polarization The solid curve shows the degree of polarization for the incoherent emitted light. The dashed curve shows the degree of polarization for the total emitted light. The degree of polarization was calculated with a dark recombination rate of the excitons of \unit[1]{ns}.}
 \label{rates}
\end{figure}

To get a more quantitative result, we show the extracted valley lifetimes in MoSe$_2$ and WSe$_2$ as a function of temperature in Fig. \ref{rates} (a). In MoSe$_2$ we find an increasing  lifetime from \unit[150]{fs} at \unit[77]{K} to \unit[400]{fs} at room temperature, which was addressed to the appearence of the energetically higher momentum-dark $(K\uparrow K'\uparrow )$ exciton states which do not contribute to the intervalley exchange coupling. At higher temperature these states get more and more populated, which leads to a quenching of the overall IEC efficiency. In WSe$_2$ we find a decrease from \unit[1.6]{ns} at \unit[77]{K} to \unit[8]{ps} at room temperature. This was ascribed to the thermalization of excitons in momentum-dark states: while at low temperatures, most excitons occupy the indirect exciton states, effectively suppressing the IEC, elevated temperatures result in a larger population of $(K\uparrow, K \uparrow)$ states, speeding up the IEC.

Experimentallly an often investigated quantity is the degree of polaraization of the emitted light $\frac{n^{\sigma +}-n^{\sigma -}}{n^{\sigma +}+n^{\sigma -}}$, which we can determine by integrating Eq. \ref{eqPL} over the time. Therefore we added a temperature independent decay constants of the excitons of \unit[1]{ns} to our numerics which can be attributed to the non-radiative recombination of excitons via defects and/or Auger processes. This decay constant is chosen in agreement with the experimentally accessible rate \cite{Zhang2015}.

The results are shown in Fig. \ref{rates} (b). In MoSe$_2$, we find very weak polarizations of the emitted incoherent light. They decrease from \unit[0.1]{\%} at \unit[77]{K} to approximately \unit[0.06]{\%} at room temperature. This behavior is a bit contradictory because one would expect an increasing degree of polarization since the valley lifetime is also increasing. The reason is the following: The degree of polarization is determined only by the excitons within the lightcone whereas we defined the valley lifetime for the total exciton density. As already discussed there is a delay between the spin relaxation of the exciton density and the emitted light at \unit[77]{K}, cf. Fig. \ref{DepolMoSe2} (a) and (b), which is determined by phonon scattering. This delay decreases drastically at room temperature due to a more efficient scattering with phonons. This results in a faster beginning of the emission of $\sigma^+$ from the bright $(K' \downarrow , K' \downarrow)$ states, which explains the observed temperature behavior. 
In the case of WSe$_2$ we find a degree of polarization of \unit[64]{\%} at \unit[77]{K} which decreases to \unit[1]{\%} at room temperature. The reason is, that a low temperatures, the valley lifetime is large, and the exciton occupation has been decayed through the non-radiative recombination before the emission from the unpumped valley sets in, cf. figure \ref{DepolWSe2} (d). At elevated temperatures, the valley lifetime, cp. figure \ref{rates} (a), decreases, in particular becomes faster compared to the non-radiative relaxation rate. This leads to an almost unpolarized emission. We want to note, that the calculated degrees of polarization depend on the choice of the non-radiative relaxation rate. Calculations for non-radiative recombination rates of \unit[500]{ps} and \unit[200]{ps} are shown in the supplementary material in Fig. 1. We find that for faster non-radiative recombination of the excitons the degree of polarization increases which would however be technologically unfavourable due to the resulting drop in quantum yield\cite{Selig2018}.

Further we investigated the impact of the coherent emission onto the valley lifetime. Since we consider a pure system without disorder in the weak excitation limit, intervalley exchange coupling does not occur for the excitonic coherence. Therefore it only emits in the polarization of the incident light. We find, that in MoSe$_2$, the degree of polarization is about \unit[64]{\%} at \unit[77]{K} and showing a decrease as a function of temperature to \unit[42]{\%} at room temperature. 
This can be addressed to the degree of coherent emitted light. At low temperatures, exciton phonon scattering is weak, so most coherent excitons decay radiatively. At elevanted temperatures, exciton phonon scattering becomes more intense, increasing the non radiative depashing of the coherent excitons. This leads to a smaller ratio of coherent emitted photons, which results in a lower polarization of the emitted light. In WSe$_2$, we observe nearly \unit[99]{\%} degree of polarization at \unit[77]{K} and \unit[89]{\%} at room temperature and find the same qualitative behavior as for MoSe$_2$ but at much larger magnitude. Under these conditions however it is technologically challenging to distinguish the sub picosecond coherent emission from the laser pulse itself. 
However, we expect that the coherent part of the emission is only accessible in experiments, if the sample is exited exactly resonant with the A exciton and the emitted light is measured in the direction of the propagation of the exciting field. 
We expect that our results for WSe$_2$ are also quantitatively applicable to WS$_2$, since here also momentum-dark exciton states ($(K,\Lambda)$ and $(K,K')$) occur\cite{Selig2016}.

Lastly, we compare our calculations for the incoherent emission with recent experimental results: 
(i) Regarding the transient regime, shortly after the optical pump, we predict an ultrafast rise of the exciton density in the pumped and unpumped valley due to the intervalley exchange coupling in both materials. This is experimentally well studied in pump probe experiments\cite{Schmidt2016,ZWang2018}. We conclude, that in the that regime, the intervalley exchange coupling is the dominant mechanism determining the intervalley dynamics, as also discussed theoretically by others before\cite{Glazov2014,Yu2014,Schmidt2016}.
(ii) Regarding times well after the optical excitation, Kerr rotation experiments revealed a biexpontential decay with the decay constants \unit[320]{ps} and \unit[5.4.]{ns} in WS$_2$ at \unit[8]{K} \cite{McCormick2018} for below gap excitation, which quantitatively matches our expectation for WSe$_2$. However, the results of reference \onlinecite{Plechinger2017} reveal, that the observed valley lifetimes in Kerr rotation drastically decrease as a function of temperature, in particular being faster than our calculated values. We conclude, that in the presence of energetically low lying momentum-dark states and the connected quenching of the IEC, other intervalley spin relaxation mechanisms\cite{Wang2014b,ZWang2018} such as pure spin flips may become faster in comparison to the IEC.
(iii) The experimentally observed polarizations of the emitted light are in the range of some ten percent for W-based TMDCs\cite{Smolenski2016}, almost no polarization of the excitonic emission was found in MoSe$_2$ at \unit[4]{K} \cite{Wang2015} which matches with our microscopic model. Regarding the dependence on the temperature, in reference \cite{Yan2015} the authors report a decrease of the polarization from about \unit[20]{\%} at \unit[77]{K} to approximately \unit[5]{\%} at \unit[200]{K} in WSe$_2$, which quantitatively matches our expectation.

\section*{Conclusion}

We have presented a microcopic theory investigating the impact of momentum-dark exciton states on the intervalley exchange coupling in monolayer transition metal dichalcogenides. We find a valley lifetime of some hundreds of fs in the absence of momentum-dark exciton states below the optically bright state (typically for Mo-based TMDCs). Contrary, we find that the valley lifetime significantly enlarges in the presence of energetically low lying momentum-dark states (typically for W-based TMDCs) which can directly be related to the relaxation of excitons into these states which are protected from the exchange interaction. While our results for the initial intervalley transfer between pumped and unpumped valley in the transient regime as well as the order of magnitude of the polarization of the emitted light are in line with recent experiments, only considering intervalley exchange coupling as spin relaxation mechanism leads to an overestimation of the valley lifetime. Therefore other spin relaxation processes such as spin flips\cite{Wang2014,ZWang2018} may become relevant at times well after the optical excitation.

\section*{Acknowledgments}
We acknowledge fruitful discussions with Dominik Christiansen (TU Berlin) and Gunnar Bergh\"auser (Chalmers). This work was funded by the Deutsche Forschungsgemeinschaft (DFG) - Projektnummer 182087777 - SFB 951 (project B12, M.S., A.K.). This project has also received funding from the European Unions Horizon 2020 research and innovation program under Grant Agreement No. 734690 (SONAR, F.K., A.K.).
The Chalmers group acknowledges financial support from the European Unions Horizon 2020 research and innovation program under grant agreement No 785219 (Graphene Flagship) as well as from the Swedish Research Council (S.B., E.M.).

\bibliographystyle{naturemag.bst}

\end{document}